# Charge state dynamics of keV ions in solids

R. Holeňák[2*], K. Vomschee[1], E. Ntemou[1,] S. Lohmann[2] and D. Primetzhofer[1,2]

[1]Materials Physics, Department of Physics and Astronomy, Uppsala University, Box 516, 751 20 Uppsala, Sweden

[2]Tandem Laboratory, Uppsala University, Box 529, 751 21 Uppsala, Sweden

**Abstract**

Fast dynamic processes between electrons in solids and a foreign atom represent a fundamental challenge for describing interactions in many-body systems and are a prerequisite for modelling materials modification. We experimentally determined the charge state distributions of slow He and Ne projectiles after transmission through thin single-crystalline silicon membranes. We found strong differences in velocity scaling and magnitude of the mean charge along different characteristic particle trajectories, providing direct insight on electron promotion and transfer processes inside the solid. Calculations of characteristic trajectories confirm the frequent spatial and ultrafast temporal accessibility of excitation channels commonly considered characteristic for large angle collisions. The commonly observed excess in energy deposition in amorphous targets compared to channelling trajectories and *ab-initio* calculations can thus be unambiguously linked to energy dissipation in frequent electron promotion as well as increased ionization density along the trajectory, driven by increased mean charge states. A quantitative comparison of energy loss and observed mean charge states further indicates complex deexcitation mechanisms at large interatomic distances masking the true equilibrium charge states along random trajectories.

Key-words: Ion transmission, Charge exchange, Molecular orbitals, Atomic collisions

*Corresponding author: radek.holenak@physics.uu.se

Accurate description of charge-transfer dynamics of energetic atomic projectiles passing through matter is mandatory for a complete understanding of radiation effects in materials modification or in biological tissues [1,2]. The detailed nature of energy-loss and charge transfer processes at energies around and below the Bragg peak has, however, been frequently challenged [3–5]. Ion transmission through 2D materials and single crystalline thin membranes was shown as an ideal geometry for accessing energy dissipation and charge-exchange dynamics in solids [6,7]. Rather straight transmission trajectories are also best comparable to the trajectories used in theoretical approaches of e.g. Time-dependent Density functional theory (TD-DFT)) [8,9] permitting to link specific processes with defined interaction distances and time scales. Notably, electron-exchange dynamics and deexcitation mechanisms were investigated using highly charged ions interacting with down to a monolayer of graphene and $MoS_2$ [10,11]. The conditions often referred to as equilibrium manifest when electron capture and loss establish constant rates, and the projectile loses memory of the initial charge state. Due to the sub-femtosecond multielectron dynamics this state is typically reached after a few monolayers. The actual processes, however, are largely inaccessible in experiments except indirectly through measurements of the aggregated energy loss. Employing self-supporting single crystalline films of several ten nanometer thickness, our recent works reported on an unprecedented increasing relative difference in the energy loss and energy-loss straggling with decreasing velocity between channeled projectiles and those experiencing comparably closer encounters with target nuclei [7,12,13]. Data strongly indicates that large quanta of energy are transferred in a series of localized events.

Today's fully atomistic first-principle calculations of electronic excitations provide a description of the electron dynamics capable of predicting the effect of changing electron densities on energy loss [14,15]. Excellent agreement was achieved with experimentally measured energy loss of light projectiles [8,16] and heavier projectiles along channeling trajectories [17]. The energy transfer along channeling trajectories is, however, often found up to an order of magnitude lower than along random trajectories and in amorphous targets [18,19], showing, that such calculations may commonly not include the processes dominating the energy loss in what is considered to be the most relevant scenario in applications. To resolve this discrepancy, models often mimic the enhanced energy loss of non-channeled trajectories by adding contributions from inner-shell electrons [20,21]. The exact nature of the contribution from inner-shell states of projectile and target atoms to electronic excitations is, however, difficult to assess also in calculations [8,9]: the relevant electrons and their whereabouts are by principle indistinguishable in the final density form and the predicted projectile charge state is an ill-defined quantity. Moreover, relevant charge-exchange process like Auger-type transitions are not yet fully incorporated in the current models [22].

In this context, other observables like emitted electrons [23–25], photons or, in particular, the projectile exit charge state [26,27] all of which carry a specific memory of only a few (if not one) last interactions, can provide access to details on charge transfer and thus energy dissipation mechanisms. Final charge state and energy following backscattering collisions have been investigated thoroughly for projectiles such as Ne [28–31]. These slow, but close collisions were shown to facilitate a sufficient level crossing of the 2p orbitals leading to electron promotion due to the formation of molecular orbitals (MO) [25,32]. Complementary studies on Auger electron emission help to identify the multitude of different excitation pathways [25].

Backscattering experiments, however, can be defined as highly specific with the interaction limited to the outermost atomic layers and small impact parameters resulting in a strong dependence of the observed charge-state distribution on penetration depth [33]. Charge-state distributions of transmitted projectiles, i.e. for much less strongly coupled nuclear and electronic energy transfers, and their trajectory dependence were previously investigated at higher velocities [19,34–36]. Multielectron processes originating in MO formation leading to projectile excitation and X-ray emission were shown to play role in establishing charge state equilibrium for projectiles with velocities near and above $v_0$ [32,37–39]. In the velocity regime $v \ll v_0$, measurements of charge state distributions are scarce with the exception of carbon studied due to its relevance as a stripper foil in time-of-flight apparatus [40–42].

In this Letter, we present experimentally obtained charge state distributions of He and Ne in the keV energy range transmitted through single-crystalline silicon membranes with a nominal thickness of 50 nm and 200 nm [43]. We reveal unprecedented differences in the exit charge-state distributions along (100)-channeling and quasi-straight pseudo-random trajectories. Despite comparable ionization potentials of He and Ne, significant differences in the exit charge-state distributions were found across the measured energy range in analogy with observed differences in the specific energy loss. The mean charge state of Ne displays a distinct velocity dependence along employed trajectories revealing that competing deexcitation mechanisms are being active during the final interaction with the exit surface. Making use of trajectory simulations we correlate the variation in the charge state distribution with the excitation of inner shell electrons in close atomic collisions. To rule out a strong impact of surface contamination all samples used in this work were pre-treated by HF-dipping. For a more detailed assessment of the sample contamination see supplementary material.

Experiments were performed with the time-of-flight medium energy ion scattering (TOF-MEIS) setup at Uppsala University [44,45]. The experiment largely follows the procedure described in [41], for the exact experiment description we refer to the supplementary material. From the measurement we extract the exit velocity of the transmitted particles and the corresponding charge state distribution.

The experimentally derived mean charge ($\langle Z \rangle = \Sigma n f_n$, where $f_n$ denotes the fraction of a specific charge) for the He and Ne projectiles transmitted through the Si membrane as a function of their exit velocity is presented in Figure 1. Differences in mean charge values are apparent between channeling and random trajectories separating towards lower and higher mean values with respect to the mean charges of He and Ne ions found after transmission through an amorphous carbon membrane in [41], [46] and [47]. Along with our data, Figure 1a features experimental data from Buck et al. [48] for $He^0$ and $He^+$ fractions derived from a subsurface signal in backscattering from a Si crystal. Figure 2 shows the distribution of individual charge states of He and Ne projectiles for channeling and random trajectories in Si selected in a narrow velocity range between 0.51 a.u. and 0.56 a.u. also marked with blue rectangles in Figure 1.

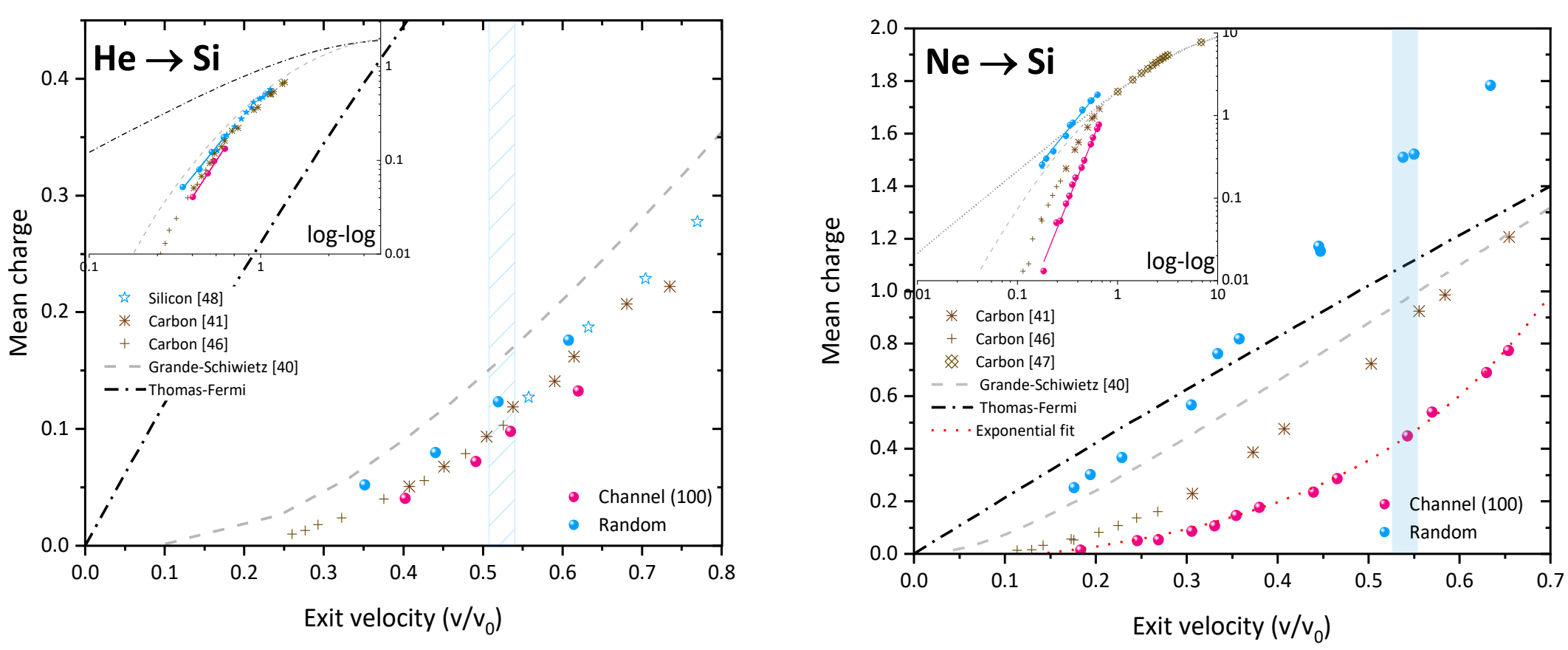

*Figure 1 Mean charge state of a) He and b) Ne ions transmitted through 50 nm and 200 nm thick single-crystalline Si membranes along with the mean charge state values for He in Si from* [48] *and He and Ne in C from* [41]*,* [46] *and* [47]*. Experimental data are compared to theoretical and semiempirical predictions by Thomas-Fermi as well as Grande-Schiwietz* [40]*, respectively. The insets indicate different velocity scaling for different trajectory types.*

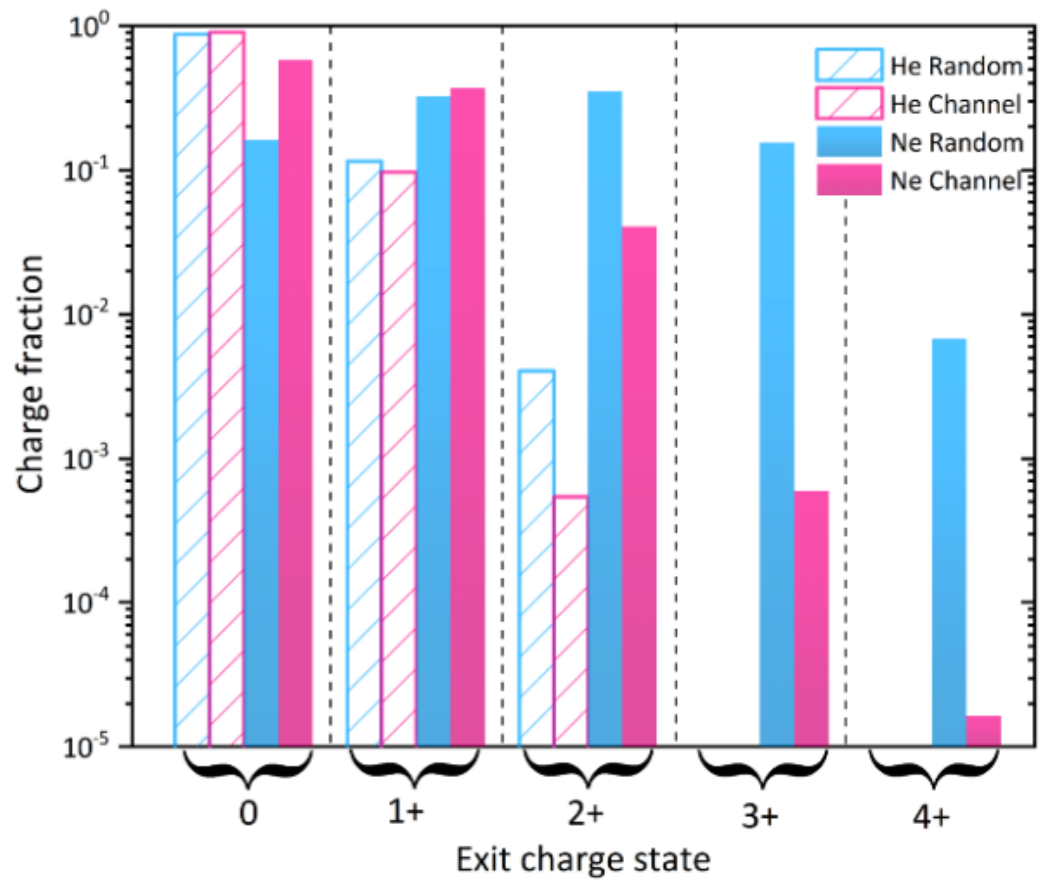

*Figure 2 Charge fractions of He and Ne at exit velocity between 0.51 - 0.56 a.u.*

*Helium*

Charge state data are well separated below and above the carbon reference all across the covered velocity range for channeling and pseudo-random trajectories, respectively. The order of the obtained mean charges from the lowest to highest reads systematically as channeling – carbon – silicon, indicating transition times through potential residual surface contaminants are significantly shorter than those necessary to establish a new equilibrium. The measured charge distribution in Figure 2 is still not exclusively determined by the number of promoted electrons inside the solid, but also by the number of captured electrons at the surface and by potentially following autoionization processes behind the solid. In contrast to $He^+$, $He^{2+}$ cannot be formed by autoionization, i.e. the nucleus leaves the sample without any electrons being captured. For coulombic interaction the probability for excitation of 2 electrons is roughly given by $p^2$ with p being the probability to excite one electron [49], in the absence of electron capture this simplifies to $p=f_1$ and $p^2=f_2$. The $He^{2+}$ fraction from the channeling measurement is, however, more than an order of magnitude below this expectation (i.e. $f_1^2>>f_2$) showing, that electron capture is dominant. The order of magnitude larger $He^{2+}$ fraction along random trajectories compared to the channeling case indicates that electron promotion due to molecular orbital interaction is more likely to promote both electrons at once.

For He, electron promotion takes place due to the antibonding interaction of the He 1s level with the target core (2p) levels accessible only in close collisions with target atoms [24,50,51]. This process requires participation of inner shell electrons and shows a non-trivial $Z_2$ dependence favoring collision partners with occupied 2p and 2s electronic shells. The order of ionized fractions from lowest to highest as channeling – carbon - silicon can be explained in this framework. Although the excitation processes are accompanied by direct energy losses (20 eV - 40 eV) [24,52], the small relative change in abundance of the higher charge fraction between the channeling and random trajectories shown in Fig. 2 would indicate that these processes are rather rare for He.

The previously reported higher energy loss along random trajectories can thus be argued to largely be a consequence of somewhat higher electron densities probed [53,54] and an increased effective charge of the projectiles [55]. Measurements of energy-loss straggling also revealed that for He, the energy is transferred predominately via direct electron-hole pair excitations of the target, irrespective of the trajectory type [13]. Mean charges observed for He are considerably lower compared to the theoretical prediction of Thomas-Fermi and display a non-linear scaling. The steep (exponential) decrease with decreasing velocity, i.e. increasing interaction time, is characteristic for neutralization processes dominated by rate-equation based Auger deexcitation at low projectile velocities. Close agreement is reached with the semiempirical prediction from Grande and Schiwietz [40] despite limited reliability of their formula in the low-velocity regime.

*Neon*

Markedly larger differences are found for mean charge states of Ne despite the comparable ionization potential of He and Ne. The separation from carbon data shows a much larger magnitude than for He. Theoretical predictions do not account for both, amplitude and scaling of measured dependencies. The mean charge for random and channeling trajectories follow different velocity dependencies. Channeling data display qualitatively similar behavior as He driven by electron loss (stripping), capture and subsequent deexcitation to the final measured charge state dominated by Auger transitions at low velocities. The significant increase in ionization along the random trajectories and the pronounced change in velocity scaling reveal a strong contribution of a charge exchange mechanism different from stripping. The higher ionization states $Ne^{+}$, $Ne^{3+}$, and $Ne^{4+}$ are orders of magnitude more abundant than in channeling geometry, while momentum transfers in collisions between electrons of ion and target exclude any direct excitation for the multiply ionized projectile. The absence of charge states higher than $Ne^{4+}$ aligns with the assumption of effective electron capture into excited states and forming the final charge by autoionization. Due to inner shell vacancies produced by MO-formation, the binding energy of otherwise unoccupied states is increased. As Ne has denser states near the Si-valence band than He, resonant electron capture to states of high main quantum number is intuitively expected to be dominant for high exit charge states [56]. The emission of an electron via subsequent autoionization is accompanied by (not necessarily complete) deexcitation of another electron. This parallel deexcitation creates a cutoff charge state dependent on the electron configuration. Assuming initial $Ne^{8+}$ with the complete 2p and 2s shells emptied, extremely efficient resonant capture of up to eight electrons occurs to excited states. In the specific example case of eight excited electrons in 3s, 3p and 3d, the detected final charge state would be $Ne^{4+}$, as for every emitted electron another electron is deexcited to 2s or 2p. However, considering higher main quantum numbers $Ne^{4+}$ can also be obtained by various configurations including only excitations of the six 2p electrons.

Employing computer simulations, we show that interaction instances probed in our transmission experiment are sufficiently small to allow for frequent effective overlap of atomic orbitals principally enabling repeated electron promotion. Recently, we have demonstrated how transmission experiments can be coupled with MC-BCA calculations to develop a better understanding of complex projectile trajectories [57]. Now, we employ these simulations to determine the distribution of interaction distances along the projectile trajectories for 33 keV He and 100 keV Ne considered in the current experiment. The final distributions of the average number of collisions per trajectory as a function of collision distance is presented in Figure 3. A pronounced difference in the distribution of collision distances is found between random and channeling trajectories, especially in context of surpassing the critical distance for a formation of MOs [28,30,58].

For illustration we display electron promotion steps corresponding to the crossing of He 1s level with the Si 2p levels and Ne 2p electron promotion along 4fσ MO formed due to the crossing with Si 2p levels. The shapes of the curves were calculated based on the formalism developed by Zinoviev [58,59] and experimental data from [28,30]. The BCA calculation for the random trajectory of 33 keV He in 50 nm Si crystal reveals that the critical distance of ~0.35 Å is reached at least a few times (approximately every 30 monolayers). For Ne, the works of Xu et al. [28] and Gordon et al. [30] reported a critical distance of ~0.6 Å in backscattering geometry, below which the promotions along 4fσ orbitals (promoting Ne 2p) become a dominant energy loss channel. For even closer collisions, promotions via other MOs (promoting Ne 2s and 1s) become possible [38,60,61]. BCA calculations for 100 keV Ne in a 50 nm silicon crystal reveal that the critical distance for MO-formation is overcome approximately every 8 monolayers. An additional relevant aspect when considering formation of MOs in the present energy regime and experimental geometry is the time spent within the critical distance: For 100 keV Ne scattering 0.7° from a Si atom (closest distance of approximately 0.4 Å) being calculated explicitly, the projectile spends only 0.097 fs below the critical distance of 0.6 Å. This time span is more than 5x shorter than in previous observations of MO formation for large angle surface scattering. For 1 keV Ne scattering under 90° from Si [30] the same closest distance of 0.4 Å is reached but already 0.842 fs are spent within the critical distance.

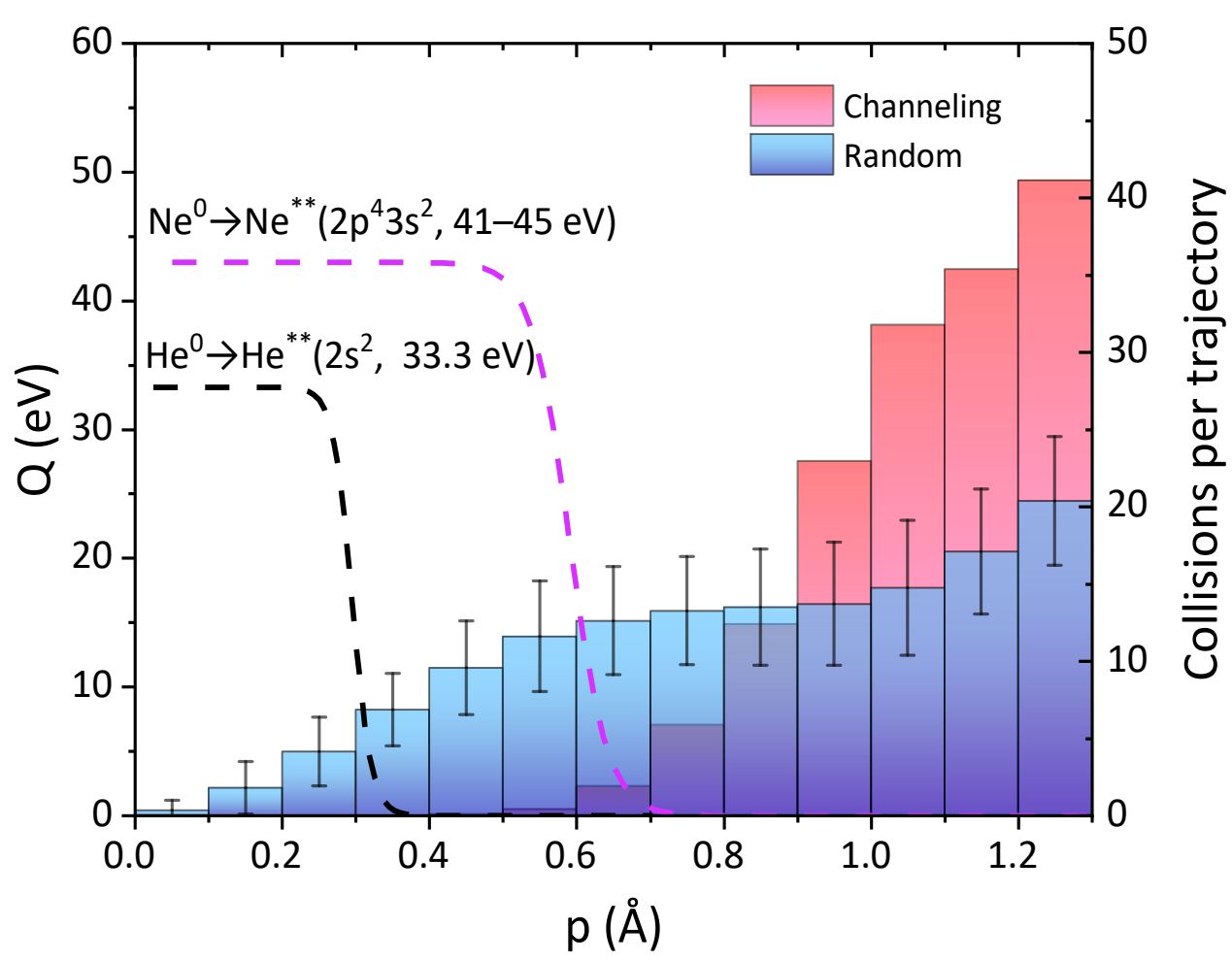

*Figure 3 Distribution of probed interaction distances per trajectory of 100 keV Ne (v = 0.43a.u.) transmitted through 50 nm silicon membrane for channelling and random orientations at <0.2° degree exit angle.*

The observed 4-5-fold difference in the mean charge reported in Figure 1 invites the question about its relation to the true effective charge state $Z_{eff}$ within the solid, if this quantity can even be defined. In an electron gas the specific energy deposition dE/dx induced by a charged projectile scales with $Z_{eff}^2$ [62] suggesting a 25-fold increase in dE/dx along random trajectories compared to channeling. Nevertheless, previous works revealed the energy loss ratio $\Delta E_{Random}/\Delta E_{Channel}$ to be 3-4 [13], in disagreement with any supralinear relation between dE/dx and $Z_{eff}$. Such discrepancy clearly suggests that at least one of the here reported charge distributions does not represent the equilibrium state. Echenique et al. [62] have compared their DFT calculations with an early channeling experiment of Eisen [18] and derived a $Z_{eff}(v/v_0 \approx 0.7) = 0.95$. Extrapolation of our channeling data to the relevant velocity regime provides close agreement with this value (see x-axis cutoff in fig. 1b)). In hyper-channeling, charge state distribution were shown to be in a close agreement with the effective charge required to provide matching angular distributions [63]. These observations suggest that the channeling charge-state distribution reflects the projectile's effective charge, consistent with an FEG-like interaction with valence and conduction electrons.

Unlike large-angle surface scattering and collisions in gas phase, a projectile traveling in the solid can undergo a number of subsequent close collisions enabling cumulative excitation of projectiles in pre-excited states. The frequency of close collisions expected from our simulations and the lifetime of formed vacancies are sufficient to enable secondary close collisions leading to even further excitation accompanied by high energy losses and to the possibility of leaving the target itself while retaining a hole in the 2s or 2p shell [31]. A series of different deexcitation pathways, such as electron capture, subsequent electron stabilization via interatomic coulombic decay and finally Auger Autoionization [11,64] will define the experimentally detected ionization state of the projectile – rendering the charge state observed for random orientation only indicative of the true nature in the solid. Based on the energy loss ratio, $Z_{eff}$ should in fact be lower than the observed mean charge. The shortcoming of the theoretical models in accurately predicting experimental data is rooted in the unaccounted influence of multi-electron processes requiring complete modelling of inner shells, meta-stable excited states and transitions of the highly excited electronic system of the projectile [40].

While the MO promotion step in Figure 3. represents an impact parameter dependent energy loss model, the exact nature of the interaction would require consideration of spatial distribution of electron densities beyond the spherical model used in BCA [65]. Nevertheless, in first approximation one can infer that for a 100 keV Ne projectile the 4fσ promotion itself could account for ~2 keV excess energy loss compared to channeling trajectories.

The here accessed magnitude of differences in the final charge-state distribution along with the variation in velocity scaling, especially for Ne, directly illustrates the crucial role of charge exchange in the interaction of keV projectiles with solids. Despite carrying strong memory of processes inside the solid, the exit mean charge does not necessarily equal the effective charge inside the solid. While our measurements match theoretical predictions for the effective charge in channeling measurements, the mean charge obtained for pseudo-random trajectories exceeds the effective charge expected from energy loss measurements. Our results proof that projectile charge and energy loss evolve dynamically along the projectile trajectory and have to be jointly considered to obtain accurate predictions. The present observations accordingly explain the observed trajectory dependence of energy loss and energy-loss straggling in one common framework. We explicitly showed that molecular orbital formation is dominant along random trajectories. In particular for heavier projectiles, we show that the specific match of the inner shells with the electronic structure of the solid can drastically change the number of available excitation and energy loss channels and via the altered mean charge state also affect the direct promotion losses. The observation of these effects in a forward scattering geometry shows that MO's form unprecedentedly fast; a time limit for MO formation might be reached at 1 a.u. projectile velocity for which extrapolation of observed mean charges in channeling and random are converging.

The results obtained here enable a thorough understanding of highly perturbative electron dynamics in solids by correlatively assessing ion charge states and energy loss moments varying atomic number and thus shell structure of projectile and target nuclei. Additional correlation with secondary particles such as photons and electrons emitted from decaying excitation states of ion and target could provide further direct information on the relevant transient electronic states.

**Acknowledgements**

Accelerator operation is supported by the Swedish Research Council VR-RFI (Contract No. 2019_00191).

# Supplementary material: Charge state dynamics of keV ions in solids

R. Holeňák[2*], K. Vomschee[1], E. Ntemou[1,] S. Lohmann[2] and D. Primetzhofer[1,2]

[1]Materials Physics, Department of Physics and Astronomy, Uppsala University, Box 516, 751 20 Uppsala, Sweden

[2]Tandem Laboratory, Uppsala University, Box 529, 751 21 Uppsala, Sweden

## Experiment

Experiments were performed with the time-of-flight medium energy ion scattering (TOF-MEIS) setup at Uppsala University [1,2]. A chopped beam consisting of pulses of singly charged ions with ~1 ns pulse length was delivered to the sample within a beam spot size smaller than 1x1 $mm^2$ and beam angular divergence better than 0.056°. The degradation of the Si-crystal is still negligible as less than 5% of the pulses repeated with 236 kHz contain an ion. Single-crystalline silicon membranes from Norcada [3] with a nominal thickness of 50 nm (±5 nm) and 200 nm (±5 nm) were positioned with the 6-axis goniometer in the center of the scattering chamber, 290 mm from the large solid-angle, position-sensitive detector. The degree of beam-crystal alignment was controlled by real-time acquisition of the 2D angular distribution and matching with trajectory calculations using the Binary Collision Approximation (BCA) code SIIMPL [4].

The membranes were chemically treated to establish a defined and reproducible surface composition. Prior to the experiment the samples were dipped in 8% HF and immediately inserted into the vacuum chamber. Transmission MEIS-recoil analysis conducted independently on a twin sample revealed a presence of residual hydrogen, carbon and oxygen: 8.4 ± 1.3 ($10^{15}$ atoms/$cm^2$), 4.0 ± 1.1 ($10^{15}$ atoms/$cm^2$) and 1.8 ± 0.6 ($10^{15}$ atoms/$cm^2$), respectively [5]. These values are very comparable to the surface composition found in the work of Bianconi et al. who adopted the same surface treatment in studying charge state distributions of projectiles scattered from the silicon surface [6]. The base pressure of the vacuum chamber was below $2*10^{-8}$ mbar.

Different charge states were discriminated with an electrostatic deflection unit positioned at a $0^{o}$ deflection angle as described in [7]. All charge states are detected simultaneously as narrow bands formed by the slit in front of the deflection plates. The charge-state distributions are obtained by separate integration of the counts within a small rectangular area in the center of each band. Stray trajectories modulated by the inhomogeneous field at the edges of the deflector are outside the evaluation window. The statistical error derived from the number of counts in the given integration window is negligible. Apparent scattering of data arises from imperfect positioning of the deflector, a change in the flight-time distance due to sample rotation and the complex angular distribution of the transmitted projectiles in random geometry induced by the blocking effect [8]. A detection limit of the charge fractions of 0.001% was achieved. The detector efficiency within the active area (60%) can be considered identical and close to unity for all atomic projectiles across the charge states. The exit velocity of the transmitted projectiles was evaluated from the central neutral charge band in the absence of an electrostatic field. The maximum final deflection angle in the sample considered in the evaluation was 0.2 degrees predefining the detectable trajectories.

## Surface Contamination

To further corroborate the origin of the observed differences in charge states and their implications on the energy-loss processes one we critically assessed the role of the target surface. The charge data presented in Figure 1 and Figure 2 of the main paper were inevitably affected by the established presence of surface contaminants, potential surface reconstruction along with variation of surface electronics states compared to bulk. To assess the reproducibility of chemical treatment and rule out a decisive effect of surface contamination on the final charge, we conducted an additional experiment for a single energy on Si membranes with three degrees of surface treatment: untreated, HF dipped, and HF dipped + annealed according to [60]. The transmission MEIS-recoil analysis revealed that after the thermal treatment the surface composition was reduced down to a single monolayer of water residues redeposited during the sample transfer and over the course of analysis. The reported results in Figure 1 show further separation between the final mean charge, i.e. more ionized projectiles along pseudo random trajectories and more neutrals along channeling. This observation further confirms the contrasting nature of the charge exchange processes along the two types of trajectories in the solid. On the one hand, surface contaminants can neutralize the projectiles exiting in a random trajectory and significantly reduce its final ionization state [61,62]. On the other hand, in the absence of close ionizing collisions with contaminants at the surface, the projectiles traversing along a channel strive further towards equilibrium charge state near zero. The cleanest sample condition was as expected reached for HF-dipping and subsequent heating. As there were only small differences between HF-dipping and subsequent heating and the more stable sample condition of only HF-dipping, we decided to record the large datasets shown in the main paper with solely HF-dipped samples.

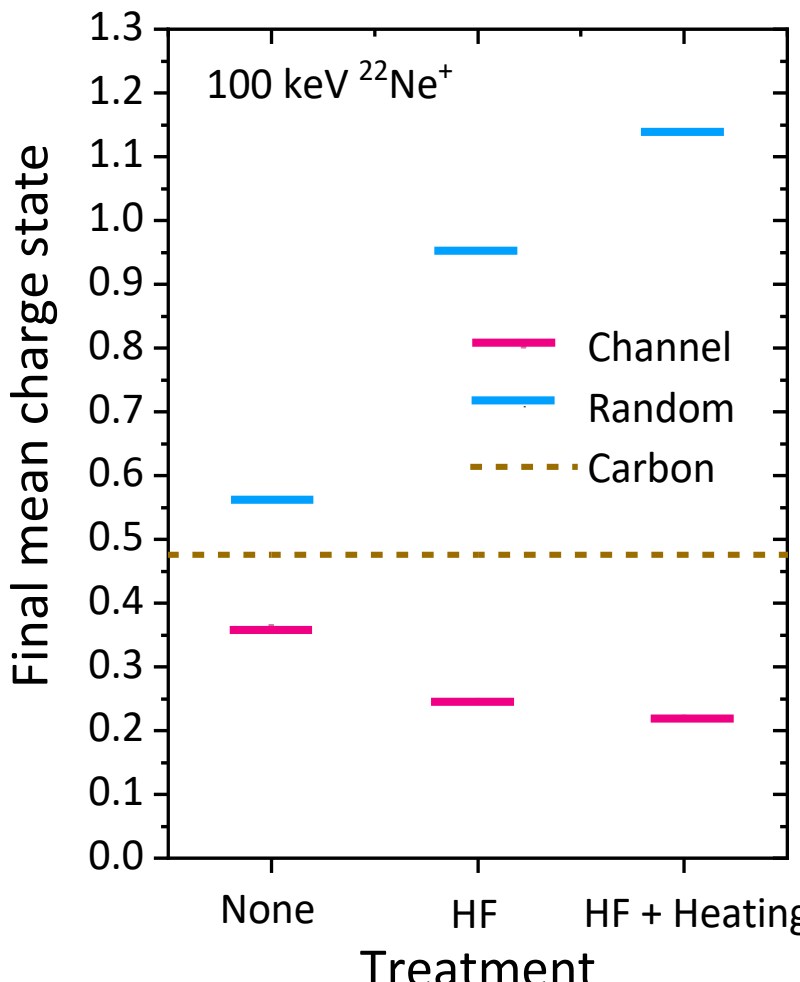

*Figure 1: The mean charge state of 100 keV Ne after transmission through 50 nm Si membrane subjected to different surface preparation routines. The increasing separation between channelling and random data from the reference carbon curve correlates with the level of oxidation and the presence of carbon contaminants on the surface of the membrane*.

## Timescales

To carefully access the timescales of the molecular orbital processes observed in the paper we performed Molecular Dynamics (MD) simulations with the code Kalypso [9]. The simulations were chosen for a typical close collision in ion transmission of 100 keV Ne and the collision investigated in 1 keV ion backscattering in reference [10].

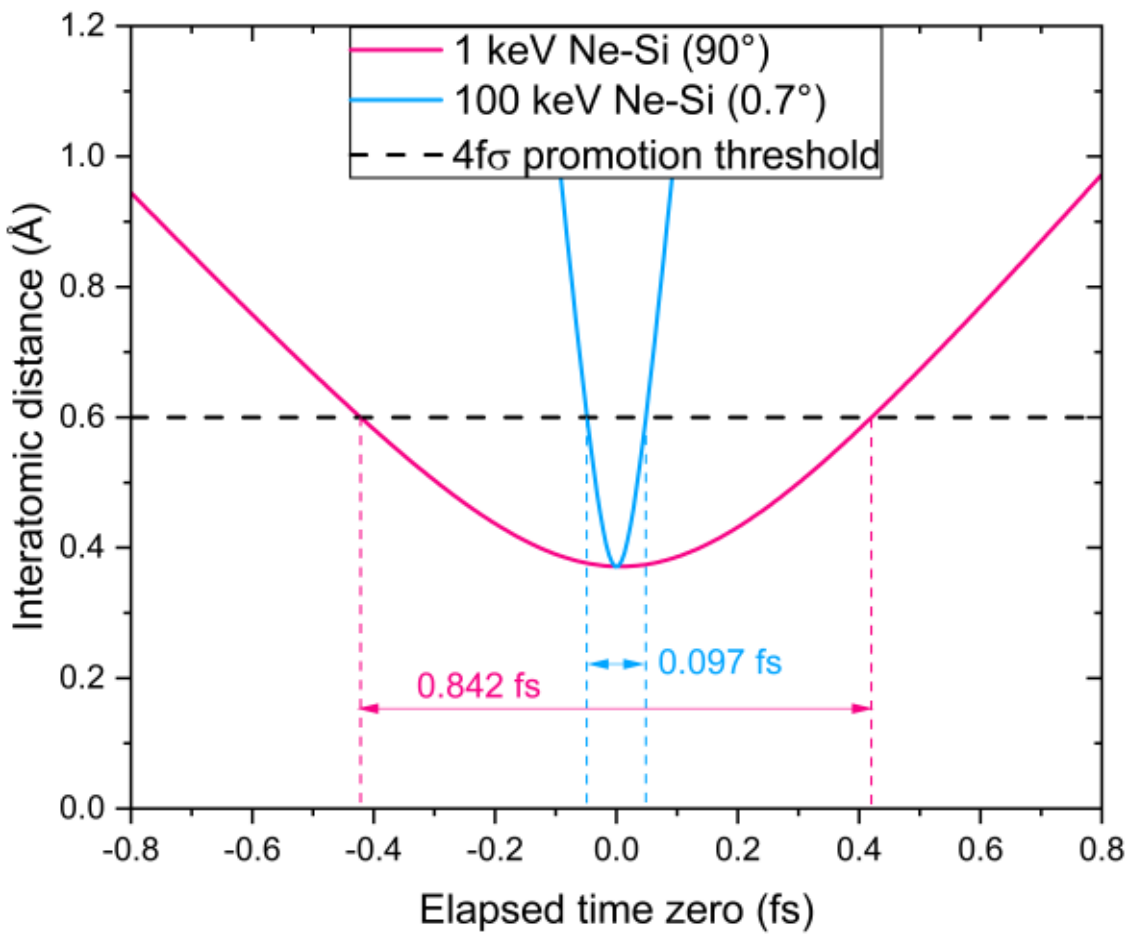

*Figure 2: MD simulation for a typical low energy surface scattering measurement and a typical close collision in our transmission experiments. The time spent below the molecular orbital promotion threshold is drastically shorter in the transmission experiment.*